\documentclass[hyperref,showpacs,prl,floats,twocolumn]{revtex4}

\usepackage{epsfig}
\usepackage{amsmath}
\usepackage{graphicx}
\usepackage{dcolumn}
\usepackage{amsmath}
\usepackage{latexsym}
\usepackage{color}

\begin{document}

\title{Disorder induced superconducting ratchet effect in nanowires}
\author{S. Poran}
\author{E Shimshoni}
\author{A. Frydman}
\address{The Department of Physics, Bar Ilan University, Ramat Gan 52900,
Israel}

\begin{abstract}
A dc voltage drop develops along amorphous indium oxide nanowires that are exposed to an ac bias source. This voltage is anti-symmetric with magnetic field and is characterized by sample specific quasi-periodic magneto-voltage oscillations. The voltage magnitude increases with decreasing temperature below $T_{C}$ but saturates at low T. As the disorder of the sample is decreased, the dc voltage is suppressed. We suggest that this rectification is a manifestation of the superconducting ratchet effect in which disorder and geometrical confinement play the role of asymmetric pinning centers. This effect demonstrates the importance of inherent inhomogeneity and vortex motion in the superconductor-insulator transition of disordered superconductors.

\end{abstract}

\pacs{74.78.Na; 73.21.Hb; 74.25.Wx; 74.81.Bd}

\date{\today}

\maketitle

Asymmetric pinning centers in superconducting films give rise to a ratchet effect in which vortices acquire a net motion in the presence of an ac driving force with time average zero. The potential exerts a counter-force with different magnitude for each direction of the driving force leading to a net velocity of vortices \cite{ratchet1,ratchet_review}. This has been demonstrated experimentally in superconductors with fabricated arrays of triangular magnetic dots \cite{triangle}, asymmetric antidots \cite{antidots,antidots2}  or Josephson junctions \cite{jj}. These systems rectify an applied ac current thus producing a dc voltage in a similar way to that of a diode. The observation of the ratchet effect has required precise fabrication of systems with a periodic and asymmetric potential array. In this letter we show that a similar effect can be achieved in geometrically confined disordered superconductors without the need to introduce artificial asymmetric pinning sites.

\begin{figure}[h]
    \centering
    \includegraphics[width=\columnwidth,keepaspectratio=true]{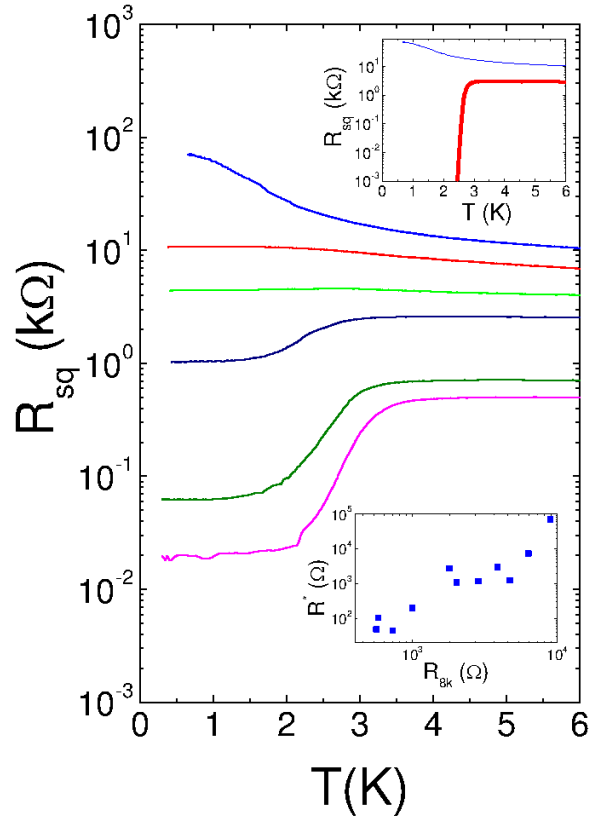}
    \vspace{-1cm}
    \caption{Resistance versus temperature for a number of InO wires with different $R_{sq}$. Top insert: R(T) of a film  with dimensions 1mm*1mm (heavy red line)  and a 1 $\mu m  \ast 50 nm \ast 30 nm$ nanowire (light blue line) having the same disorder. Bottom inset: residual resistance $R^*$ as a function of the normal state sheet resistance (color online).}
\end{figure}

Amorphous indium oxide (a-InO) films undergo a superconductor insulator transition (SIT) as a function of disorder that can be controlled by low temperature annealing \cite{shahar1}. Despite the fact that the films are amorphous and are relatively structurally uniform \cite{hebbard} there have been a number of experimental observations that indicate that the electronic structure is inherently inhomogeneous on rather large length-scales ($\leq 1\mu m)$. Two examples for this are an SIT as a function of the length in a set of a-InO short samples \cite{kowal} and local STM spectroscopy measurements showing spatial variations of the order parameter in InO films \cite{sacepe}. Indeed, electronic "granularity" has been predicted to emerge in homogeneous disordered superconductors by a number of groups \cite{efrat,nandini,dubi,imry}. Such inhomogeneity can be expected to have especially large impact in low dimensional systems in which at least one of the spatial extents is smaller than the scale of granularity.

The samples used in this study were lithographically defined InO nanowires, 30 nm thick, 40-90 nm wide and 1-5 $\mu m$ long. The InO was e-gun evaporated in a vacuum chamber with base pressure down to $1 e^{-7}$ mbar in which partial oxygen pressure up to $1 e^{-4}$ mbar was introduced to produce samples with different initial disorder. The disorder was further reduced by vacuum annealing within the measurement probe at $T=80 ^\circ C$ causing a continuous decrease of the wire resistance. The normal state resistance of the wires at $T=10\,K$ ranged between $1 \, M\Omega$ to $10 \, k\Omega$. We have studied 14 such wires at different degrees of disorder, all showing the same qualitative behavior described below. Similar wires produced by a different technique were studied by Johansson {\it et al.} \cite{shahar2}. These wires exhibited low temperature reproducible magneto-resistance oscillations. In this work we focus on the response of such wires to an applied ac bias.

Throughout the experiment, an ambient radiation of 15.5 MHz, 0.02 mW was imposed on the wires. Under these conditions even the most ordered wires did not reach a zero resistance state at low T. Figure 1 shows the resistance versus temperature for a series of wires having different degrees of disorder which are characterized by the sheet resistance $R_{sq}$. It is seen that the wires undergo a disorder driven SIT. However, even on the superconducting side of the transition the wire resistance saturates at a low temperature residual resistance, $R^{*}$, which increases with increasing disorder (or $R_{sq}$, see inset). Interestingly, also on the insulating side the resistance seems to be approaching a finite $R^{*}$ rather than increasing exponentially with lowering T. 2D films prepared with the same disorder as these wires always showed full superconductivity even in the presence of an induced ac bias as demonstrated in the inset of Fig. 1.

The observed finite resistance is not surprising. The superconducting coherence length, $\xi$, in disordered indium oxide films is 10-30 nm \cite{shahar2} which is not much smaller than the width of our wires. One can expect the imposed radiation to induce phase slips or vortex anti-vortex motion which would generate dissipation in the wire (though the temperature dependence needs clarification). A feature that is much less expected is the appearance of a \emph{dc} voltage along the wire which is anti-symmetric with magnetic field. Figure 2 depicts a magneto-voltage (V(H)) curve of a $1 \mu m$ long wire. This voltage was measured by a Keithley 2182A nanovoltmeter connected to the two ends of the wire without any driving current. It is seen that the voltage increased rapidly up to a field $H_m$ at which point it reached a maximal value $V_m$. For $H>H_m$ the voltage decreased gradually. This qualitative behavior, which was seen for all measured wires, is very typical of vortex motion in the presence of a driving bias such as that reported in Nernst effect measurements performed on a-InO films \cite{aubin1,aubin2}. Figure 2 also shows that, for our wires, additional fine-structure was superimposed on the overall V(H) curve. These features were sample dependent but were very reproducible for a single wire even after repeated thermal cycling and showed nearly perfect antisymmetric behavior, $V(H)=-V(-H)$. Fourier transforming the V(H) curves of our wires (after subtracting the large background) yields a dominant quasi-period (see inset of Fig. 2) which corresponds to a magnetic field of 0.6-1.5T. This is consistent with a magnetic flux quantum penetrating a loop of diameter 20-50 nm which is close to the width of our wires. Hence, it seems reasonable to attribute each oscillation to an additional vortex penetrating the wire.

A rectification of an ac driving force which is anti-symmetric with H is a fingerprint of the ratchet effect of vortex motion. This effect requires a periodic array of asymmetric pinning centers. Such an array is not intentionally introduced to our wires. One should recall, however, that the wires are highly disordered and may be characterized by inherent inhomogeneity. Hence, they may include regions of normal metal or weak superconductivity that may act as pinning centers for vortices. The geometry of the wire, in which the width is not much larger than the scale of inhomogeneity, gives rise to natural asymmetry of pinning centers that are closer to one edge of the wire than the other. Hence, the interplay between disorder and the wire geometry makes it easier for a vortex to cross the barrier to one side than to the other, thus producing a unique ratchet effect.

\begin{figure}
    \centering
    \includegraphics[width=\columnwidth,keepaspectratio=true]{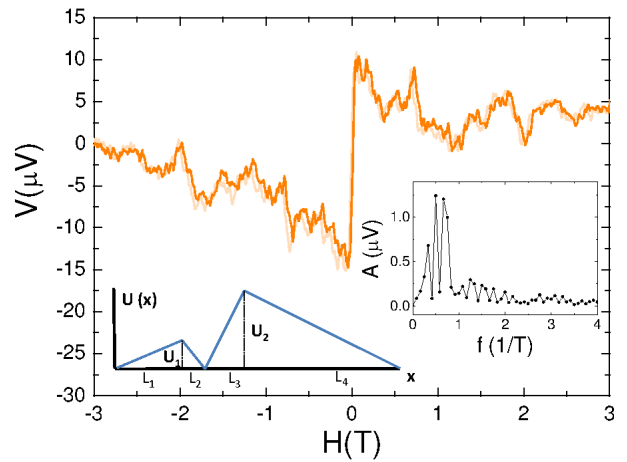}
    \caption {dc voltage as a function of magnetic field for a $1 \mu m \ast 30 nm \ast 50 nm$ InO wire. The two curves are a trace and a retrace showing that the fine-structure is very reproducible. The right inset shows the fourier transform of the curve after subtracting the main feature background. The left inset is a schematic illustration of  the pinning potential U(x) along the superconductor's cross-section.}
\end{figure}

A simple model for the pinning profile across the wire, U(x), due to a single pinning site placed asymmetrically with respect to the wire width is illustrated in the inset of Fig. 2. Here the superconducting regions within the wire act as barriers for vortices and the normal insulators outside the wire are vortex reservoirs. A vortex in the potential well is subject to an ac driving force with amplitude $f_d=j_{ac}\Phi_0d/c$ where $j_{ac}$ is the ac current density, $\Phi_0= 2\cdot10^{-7} Gcm^2$ the flux quantum and d the film thickness. For small amplitudes of driving force, $f_d\ll \frac{dU}{dL}$, L being the barrier length, the dc voltage, V, is proportional to the difference between the probabilities to cross over the barriers:
\begin{equation}
\label{eq_excitation}
V \propto [e^{-\frac{U_1-f_dL_2}{k_BT}}-e^{-\frac{U_2-f_dL_3}{k_BT}}]
\end{equation}
where $U_1$ and $U_2$ are the potential barriers for a vortex to escape the wire in either direction.

Equation 1 predicts that V should increase with increasing disorder (which would increase the relevant potential barriers) and with decreasing temperature. Both these trends were observed in our wires. Though the voltage magnitude varies considerably from sample to sample making comparison between different samples difficult, in a single wire the dc voltage showed monotonous decrease upon reducing the disorder. This is demonstrated in Fig. 3a which depicts V(H) of a single $1 \mu m$ long nanowire for three annealing stages. It is seen that $V_m$ decreases as the sample is driven deep into the superconducting state. At the same time, $H_m$ is also pushed to lower fields.

\begin{figure}
    \centering
    \includegraphics[width=\columnwidth,keepaspectratio=true]{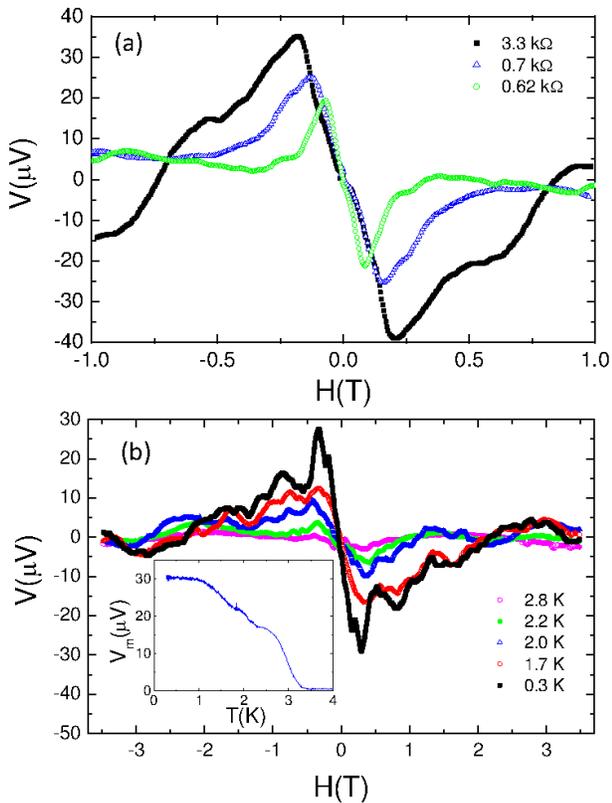}
    \caption{Top: V(H) curves for a $1 \mu m \ast 30 nm \ast 50 nm$ InO wire for three stages of disorder, achieved by thermal annealing. Bottom: V(H) curves for a different $1 \mu m \ast 30 nm \ast 50 nm$ InO wire measured at various temperatures. The inset shows the maximal voltage, $V_m$, as a function of temperature.}
\end{figure}

The temperature dependence of V(H) for a typical wire is shown in Fig. 3b. It is seen that below $T_C$ the curve increases in amplitude with lowering T as expected from Eq. 1. However, below a characteristic temperature, $\tilde{T}$, the voltage saturates and becomes temperature independent as seen in the inset of Fig 3b. $\tilde{T}$ is found to be between 1.3K and 2K in all our wires and is very close to the temperature at which the resistance reaches the saturation value $R^*$ (Fig. 1). The saturation of V(T) may be a result of a crossover from thermal excitation to quantum tunneling of vortices across the barriers. Such interpretation clearly requires a detailed theoretical treatment and further experimental investigation.

When the driving force is increased so that it approaches the force exerted by the pinning potential the barrier is suppressed and the exponential activation form of Eq. 1 does not apply. The vortices become free and their motion can be described by a viscous flow \cite{ratchet1}. For each pinning center, having a potential profile such as that illustrated in the inset of Fig 2, the velocity due to the viscous flow which is proportional to the rectified voltage can be shown to be given by:

\begin{eqnarray}
\label{voltage}
v=\left\{
\begin{array}{l}
    0 \hspace{4.4cm}{\rm if}\,\,\, f_d<f_2 \\ \\
    \frac{1}{2\mu}\frac{(f_d+f_1)(f_d-f_2)}{f_d+f_1-f_2} \, \hspace{2cm} {\rm if}\,\,\, f_2<f_d<f_3 \\ \\
    \frac{1}{2\mu}(\frac{(f_d+f_1)(f_3-f_2)f_1}{(f_d+f_1)(f_d+f_1-f_2-f_3)+f_2f_3} \, \,\, {\rm if}\,\,\, f_3<f_d\
\end{array}
\right.
\end{eqnarray}
where $f_2=\frac{U_1}{L_2}$, $f_3=\frac{U_2}{L_3}$, $f_1=\frac{U_1}{L_1}=\frac{U_2}{L_4}$ are the forces generated by the pinning potential and $\eta$ is the viscous drag
coefficient for vortex motion.

Equation 2 describes a non monotonic dependence of the vortex velocity on the magnitude of the ac driving current. The reason for this is that when the driving force is very low, vortex motion is limited for both bias orientations and when it is very high it will generate equivalent vortex motion in both directions. Thus, the voltage reaches a maximum at intermediate force amplitude, $f_d=f_3$, and is very small for both high and low $f_d$ amplitudes. In our wires the voltage may be governed by a number of pinning sites, each characterized by a different potential profile and contributing differently to the total voltage drop. In this case the dependence of voltage on applied ac current magnitude can be nontrivial and include a number of maxima. Figure 4 shows the voltage across a  cvwire as a function of the applied ac bias power, P. A number of points should be noted. First, there is a threshold ac power, $P_{min}$ below which no voltage is detected. $P_{min}$ is magnetic field dependent. Second, the voltage is non monotonic with ac power and shows a series of peaks. Finally, depending on the magnetic field, the voltage can be either positive or negative. At certain magnetic fields the voltage may change sign as a function of bias power thus manifesting a "ratchet reversal effect" \cite{triangle}. This behavior reflects the summation of contribution from a number of vortex pinning sites, each one having different asymmetry and hence a specific ac power and magnetic field dependence of voltage magnitude and sign.

The longer the wire, the more pinning centers can be expected to contribute. The total voltage is given by $V=\sum V_{i}$ where $V_i$, the contribution of pinning site i, is random in size and can be either positive or negative. V is the sum of random numbers and is expected to increase in magnitude like $V \propto\sqrt{N}$ (where N is the number of sites generating random voltage) in a similar way to a one dimensional random walk of random step size. Increasing the wire length should lead to a richer structure of $V(f_d)$ and to larger V. We have measured two $ 5 \mu m$ and 12 $ 1 \mu m$ long wires. The long wires yielded voltages that were more than twice larger than the average voltage measured in shorter wires having equivalent disorder. The longer wires also showed richer and more prominent fine structure as a function of magnetic field and ac bias power. An opposite trend can be expected with increasing wire width. The ratchet effect should be suppressed when the width exceeds the scale of inhomogeneity.  Indeed, we did not detect any dc voltage for wires having widths of several $\mu m$. A detailed study of the dc voltage dependence on wire length and width will be presented and discussed in the future \cite{future}.

\begin{figure}
    \centering
    \includegraphics[width=\columnwidth,keepaspectratio=true]{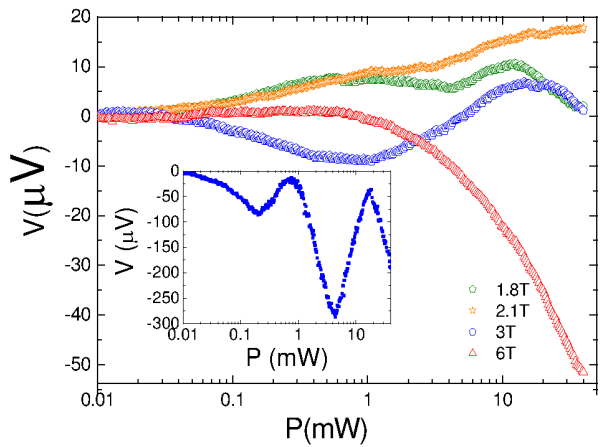}
    \caption{dc voltage as a function of ac power for an InO wire measured at different magnetic fields which correspond to peaks in the V(H) curve. Inset: Voltage as a function of power at H=2.4T for a $5 \mu m$ long sample. Note that the curve is composed of a number of voltage maxima and minima.}
\end{figure}

In summary, this work suggests a new type of superconducting ratchet effect in which there is no need for a periodic array of artificial asymmetric pinning sites. In our wires disorder played a significant role but it should be possible to engineer a ratchet device for example from a narrow bridge of a clean superconductor with one antidot placed closer to one of the edges. Increasing the number of the pinning centers along the bridge would increase the voltage making it an effective detector for radiation. The results of this work also provide a clear illustration of the significance of inherent electron inhomogeneity in driving the SIT of these materials. In particular they demonstrate that vortices are instrumental in the processes that lead to dissipation and electric resistance in disordered superconductors.

We are grateful for useful discussions with C. Bolech, M. Mueller, Z. Ovadyahu, G. Refael, N. Shah and D. Shahar. This research was supported by the US Israel binational fund (grant No. 2008299). E.S. acknowledges support from the Israel Science Foundation (grant no. 599/10) and from the Israeli ministry of science and technology (grant 3-5792).

\end{document}